\newif\ifIEEE
\IEEEtrue
\newif\ifACM
\ACMfalse

\newif\ifINFLATE
\INFLATEtrue
\ifIEEE
\documentclass[10pt,conference]{IEEEtran}
\fi

\ifACM
\documentclass{sig-alternate-05-2015}
\fi

\usepackage{graphicx}

\ifACM
\setcopyright{acmcopyright}
\CopyrightYear{2018}
\doi{http://dx.doi.org/xx.xxxx/xxxxxxx.xxxxxxx}
\isbn{978-1-4503-3739-7/16/04}
\conferenceinfo{CryBlock 2018,}{ June 15, 2018, Munich, Germany}
\acmPrice{\$15.00}

\fi

\title{Towards Application Portability on Blockchains}
\ifIEEE
\author{%
\IEEEauthorblockN{Kazuyuki Shudo}
\IEEEauthorblockA{
Tokyo Institute of Technology\\
Tokyo, Japan
\ifINFLATE
\\Email: shudo@c.titech.ac.jp
\fi
}
\ifINFLATE
\and
\IEEEauthorblockN{Reiki Kanda}
\IEEEauthorblockA{
Tokyo Institute of Technology\\
Tokyo, Japan
\\Email: kanda.r.aa@m.titech.ac.jp
}
\fi
\and
\IEEEauthorblockN{Kenji Saito}
\IEEEauthorblockA{
Keio University\\
Kanagawa, Japan
\ifINFLATE
\\Email: ks91@sfc.wide.ad.jp
\fi
}
}
\fi

\ifACM
\numberofauthors{2}
\author{
\alignauthor
Kazuyuki Shudo\\
    \affaddr{Tokyo Institute of Technology}
\alignauthor
Kenji Saito\\
    \affaddr{Keio University}
}
\fi

\begin{document}

% To allow figures and tables to occupy a large area
\renewcommand{\topfraction}{1.0}
\renewcommand{\dbltopfraction}{1.0}
\renewcommand{\textfraction}{0.0}
\setcounter{topnumber}{3}
\setcounter{bottomnumber}{3}
\setcounter{totalnumber}{5}

\maketitle

\begin{abstract}
We discuss the issue of what we call {\em incentive mismatch}, a fundamental problem with public blockchains supported by economic incentives.
This is an open problem, but one potential solution is to make application portable.
Portability is desirable for applications on private blockchains.
%It is not even clear to be able to define a common API for various blockchain middlewares, but it is possible to improve portability by reducing dependency on a blockchain.
Then, we present examples of middleware designs that enable application portability and, in particular, support migration between blockchains.
\end{abstract}

\ifIEEE
\fi

\ifACM
\fi

\section{Introduction}

%Test \cite{url:Ethereum,url:HyperledgerFabric,Saito2017:BBc1,KaSh17e,GKWG16,GaKL15,DeWa13,CaLi02,Naka08}.

\ifINFLATE
Blockchain is extending its field of applications, not limited to cryptocurrency, due to its ability to provide us a kind of trust even with no centralized entity such as a government as shown in Figure \ref{fig:bc-overview}.
\fi

{\em Incentive mismatch} is a fundamental problem with public blockchains.
The incentive for nodes to support a blockchain is economic, i.e., gaining coins, and is different from the incentives for blockchain applications.
An application cannot continue working if its underlying blockchain collapses due to the economic motivation disappearing.
%If possible, it is better for them to match.
%But it is not so trivial to align them and it is still an open problem.
It is non-trivial to align the blockchain node and application incentives, and this is still an open problem.

Making applications portable is one potential solution to protecting them against collapsing along with their underlying blockchains.
Portability is also desirable for applications on private blockchains.
In fact, minimizing the dependence of applications on their underlying middleware is a well-known best practice.

However, current blockchain middlewares provide their own application programming interfaces (APIs), making an application written for one middleware difficult to port to another.
It is not even clear that it would always be possible to design common API functions, because each middleware uses its own abstractions, such as Ethereum's Solidity language.

In this paper, we introduce the incentive mismatch problem and discuss application portability as a potential solution.
Then, we present software architectures and techniques for enabling application migration between blockchains.

% Figure 1
\ifINFLATE
\begin{figure}[t]
\centering
\includegraphics[width=.48\textwidth,clip]{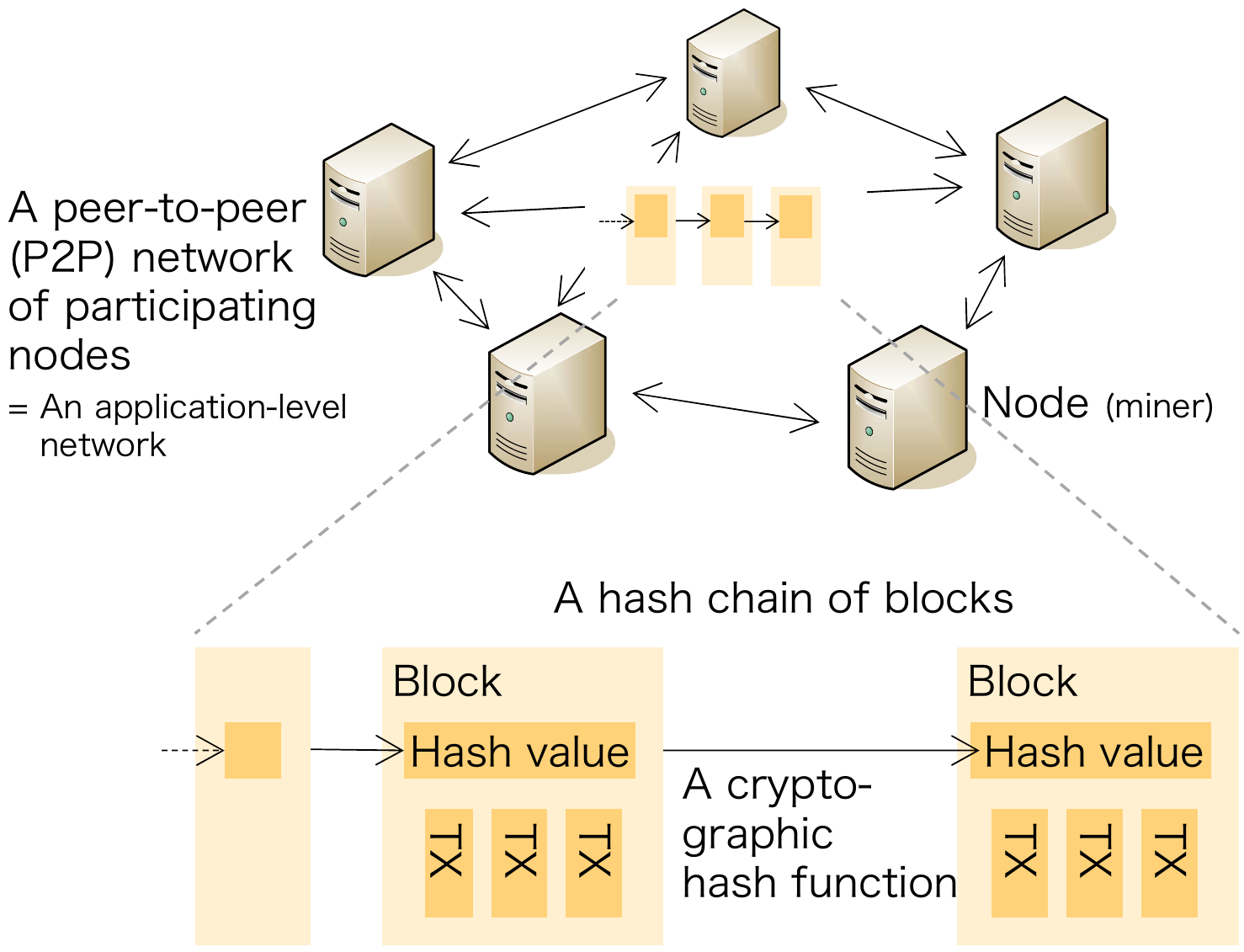}
  \caption{A blockchain depicted as a distributed system.}
  \label{fig:bc-overview}
\end{figure}
\fi

\section{Benefits of application portability}

Portability is a desirable property for applications on both public and private blockchains, although the reasons are somewhat different.
%But blockchain middlewares provide their own APIs and applications are hardly portable.

\subsection{Applications on public blockchains}
\label{sec:motiv-pub}

Public blockchains have a fundamental problem we call {\em incentive mismatch}:
blockchain nodes and applications do not share a common incentive.
Public blockchain transactions are confirmed by Proof of Work \cite{DwNa92}, and derivatives such as Proof of Stake \cite{Quan11} and Proof of Elapsed Time \cite{url:HyperledgerSawtooth}, that are based on the nodes' economic incentives.
The nodes simply try to gain coins and have no direct incentive to support particular applications.

If a public blockchain cannot provide sufficient economic incentives to its supporting nodes, it loses the ability to confirm transactions securely.
For example, a fall in coin prices might lead to some of its nodes defecting, and reducing its ability to confirm transactions.
Blockchains with fewer supporting nodes are also more vulnerable to attacks such as majority (51\%) attacks and eclipse attacks \cite{GRKC15,HKZG15}.
For example, in May 2018, attacks against public blockchains supporting cryptocurrencies succeeded in a row.
In case of Monacoin, attackers voided over 20 blocks that had previously been confirmed by a block withholding attack.
In case of Bitcoin Gold, attacker voided 22 or more blocks and succeeded double-spending of coin.
This means that blockchain applications have no control over the confirmation ability they are based on and, hence, they can collapse due to economic circumstances.

\ifINFLATE
%
% Figure 2
\begin{figure*}[t]
\centering
\includegraphics[width=.65\textwidth,clip]{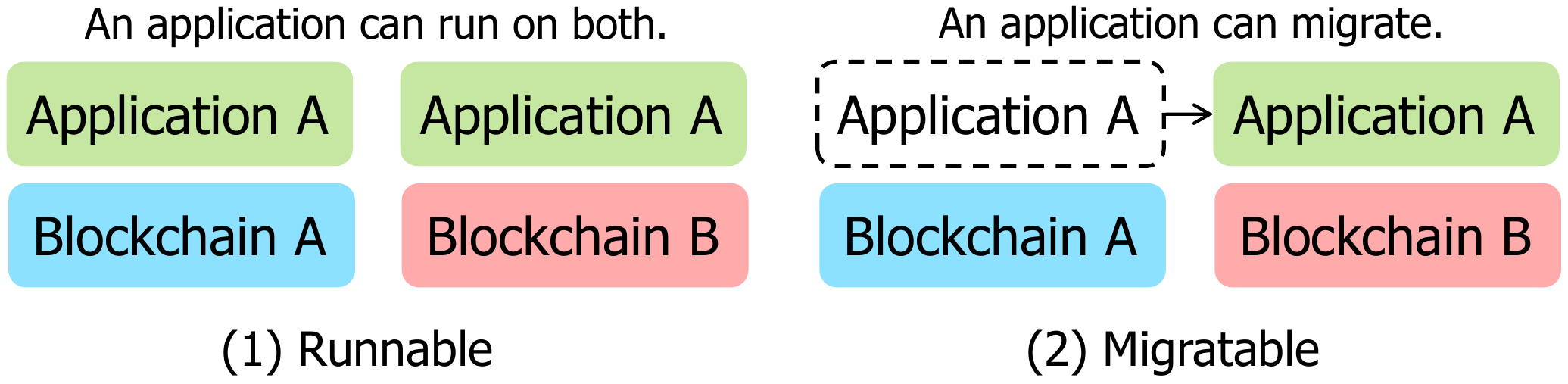}
  \caption{Two application portability levels.}
  \label{fig:portability-levels}
\end{figure*}
\fi

Is it possible to align the incentives of the nodes supporting a blockchain and the applications on it?
%It depends on an application.
%
The currency system that supports a public blockchain is itself a blockchain application, because it utilizes the blockchain's confirmation ability.
Their incentives match by design; however, this is a special case.
The incentives also match if the same entity both deploys and operates the blockchain nodes and runs the application; however, in that case it essentially becomes a private blockchain.

Otherwise, public blockchains and their applications generally have no common incentive.
Whether or not it is possible to align their incentives, i.e., design a mechanism that enables applications to ensure they retain this confirmation ability, is an open problem.
However, making application portable is one potential solution to this problem:
if an application is portable (i.e., migratable), we can simply migrate it to another blockchain if the current one collapses (Section \ref{sec:portability}).

\subsection{Applications on private blockchains}
\label{sec:motiv-priv}

Portability is also a desirable property for applications on private blockchains.
If it is difficult to port a given application, it may suffer from so-called vendor lock-in, preventing, for example, the application and its users benefiting from a different and better middleware platform.
For example, an application and its users cannot benefit from a better another middleware.
At worst, this means the application will die along with its underlying middleware if that middleware becomes too old and stale to use.

\section{Application portability}
\label{sec:portability}

% todo: \subsection{Portability levels}

\ifINFLATE
\else
%
% Figure 2
\begin{figure}[t]
\centering
\includegraphics[width=.48\textwidth,clip]{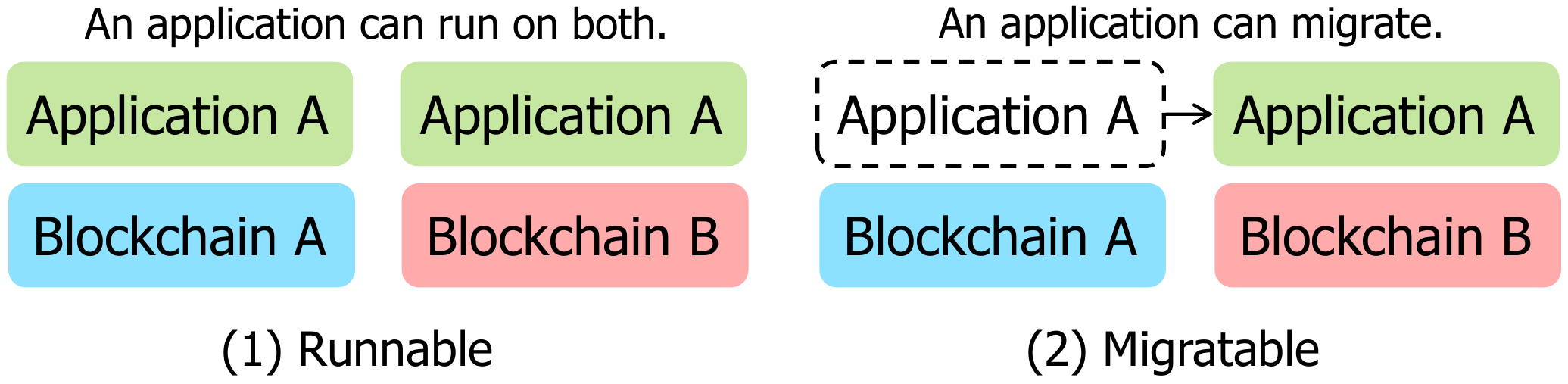}
%
%  \vspace{-2.0ex}
%
  \caption{Two application portability levels.}
%
%  \vspace{-3.2ex}
%
  \label{fig:portability-levels}
\end{figure}
\fi

There are several different levels of application portability.
As Figure \ref{fig:portability-levels} shows, here, we focus on the following two levels.
\begin{enumerate}
\item{{\em Runnable} :}\quad The application can run on different blockchains, but cannot migrate from one to another.
\item{{\em Migratable} :}\quad Both the application and its data can migrate from one blockchain to another.
\end{enumerate}

Using a common blockchain API enables {\em runnable} portability.
Applications are also clearly portable between different blockchains running the same middleware.
However, it is less clear whether we can design common API functions for use by different middlewares, because they often adopt different abstractions, for example adopting a directed acyclic graph (DAG) instead of a hash chain of blocks. In addition, each middleware has its own smart contract mechanism, such as Ethereum's Solidity language.
However, applications can still be portable if we limit the blockchain functions they use and provide a common API for this (more limited) set of functions.
For example, all blockchains should support storing a hashed value together with its time stamp, and we can provide a common API for that function.

If an application is portable at the {\em runnable} level, it can only be migrated to another blockchain by restarting it in its initial state and accepting the loss of all the accumulated data.
Because proof of data existence of and verification of state changes are fundamental blockchain features, migration without the logs needed to enable those features (Section \ref{sec:mig-data}) is unlikely to be useful.
In contrast, the {\em migratable} portability level enables applications to survive both the loss of private blockchain middleware (Section \ref{sec:motiv-priv}) and the collapse of an underlying public blockchain (Section \ref{sec:motiv-pub}).

\section{Migrating applications between blockchains}

In this section, we discuss a method of enabling application migration between blockchains, and present preliminary middleware designs based on it.

\subsection{Data to be migrated}
\label{sec:mig-data}

What data should be migrated between blockchains?
When an application that does not depend on a blockchain is migrated between middlewares, it is generally sufficient to migrate its current state, for example, records in relational databases.
However, one of the strongest reasons for using blockchains is that they can prove data existed at a given time in the past and verify state (data) changes.
In this case, migrating just the application's current state is not enough; the logs enabling these proof and verification processes must also be migrated.
Thus, we need to migrate the following two types of data.
\begin{enumerate}
\item {\em Current state} of the application
\item {\em Logs} -- metadata of the states that describe state changes and their time stamps.
\end{enumerate}
In currency applications such as Bitcoin, for example, the former data are account balances, although, in Bitcoin, these are not explicitly recorded and can instead be calculated by adding up related transactions.
The latter data are transactions, that describe balance changes and their time stamps.

\ifINFLATE
%
% Figure 3
\begin{figure*}[t]
  \centering
  \includegraphics[width=.70\textwidth,clip]{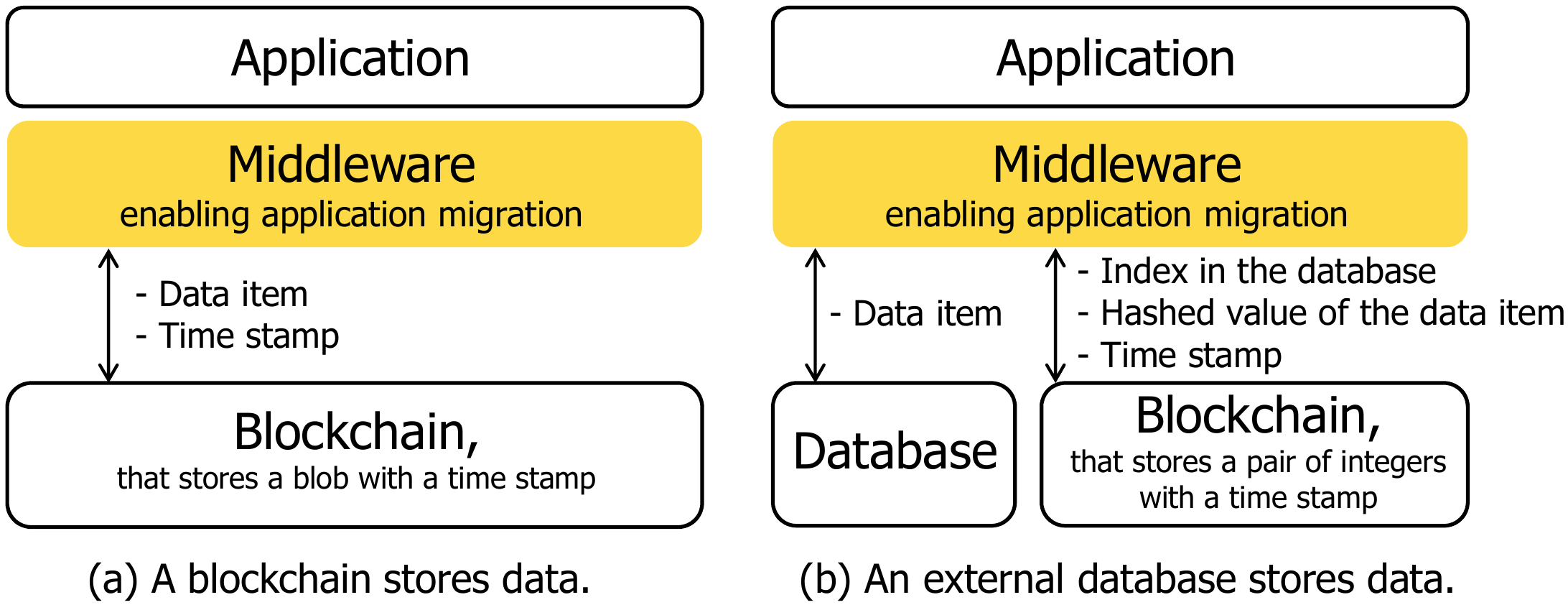}
  \caption{A software architecture that improves application portability by limiting dependency on a blockchain.}
  \label{fig:arch}
\end{figure*}
\fi

\subsection{Middleware design that enables migration}
\label{sec:arch}

We adopt the following two principles to design migration-friendly middleware.
\begin{itemize}
\item Minimize dependence on specific blockchain middleware.\\
The only requirement for our middleware design is to support the storage of simple data items, such as numbers or byte sequences, together with associated time stamps.
\item Do not expect to be able to retain trust in the original (source) blockchain.\\
One of the main reasons for an application to leave its current blockchain is imminent collapse of the blockchain, for example, due to the loss of too many nodes (Section \ref{sec:motiv-pub}).
In such a situation, we cannot continue to rely on the previous blockchain.
\end{itemize}

\ifINFLATE
\else
%
% Figure 3
\begin{figure}[t]
  \centering
  \includegraphics[width=.49\textwidth,clip]{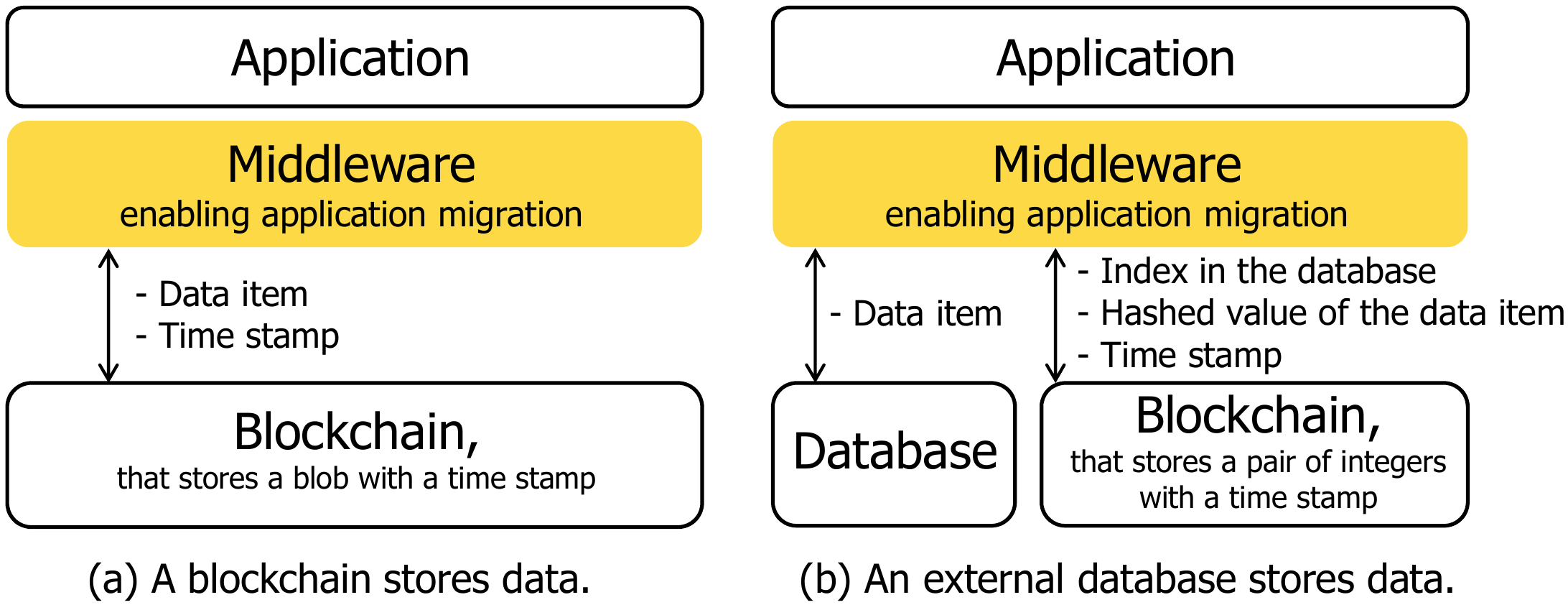}
%
%  \vspace{-1ex}
%
  \caption{A software architecture that improves application portability by limiting dependency on a blockchain.}
%
%  \vspace{-2ex}
%
  \label{fig:arch}
\end{figure}
\fi

Figure \ref{fig:arch} shows two candidates for middleware designs that enable migration between blockchains.
To minimize the dependences discussed above, these only expect the underlying blockchain to support recording a set of numbers and a time stamp with its indexing key (e.g. account ID such as Bitcoin address).
At most, we expect a time-stamped byte sequence.
We will investigate enabling migration while supporting more features than this in future work.

In the design shown in Figure \ref{fig:arch} (b), we have chosen to store the data in an external database, not in the blockchain, although the blockchain still contains information that must be migrated.
The Beyond Blockchain One (BBc-1) blockchain middleware \cite{Saito2017:BBc1} adopts such a design.
This approach reduces the blockchain size, which is advantageous because the blockchain is copied to all nodes and thus occupies storage space on all of them.
In addition, if an application does not require certain state (data) changes to be verified, we can update the data simply by overwriting it in the database, further reducing the storage requirements.

With the design shown in Figure \ref{fig:arch} (b), we must also ensure the database is sufficiently fault-tolerant to achieve enough data availability.
We can achieve this by utilizing replication or erasure coding \cite{FLPS10}, which are supported by most distributed databases, and adjusting the fault tolerance level by configuring the number of replicas or the erasure coding parameters.
This flexibility is an advantage over the design shown in Figure \ref{fig:arch} (a) that simply copying everything to all the blockchain nodes.
Currently, such highly-available databases are provided as public cloud services and we can even store data in multiple databases instances across different cloud providers.

In the design shown in Figure \ref{fig:arch} (b), if an application does not require verification of state (data) changes, we can overwrite data in the database when they are updated.
It reduces the amount of storage occupation further.

\ifINFLATE
%
% Figure 4
\begin{figure}[t]
  \centering
  \includegraphics[width=.48\textwidth,clip]{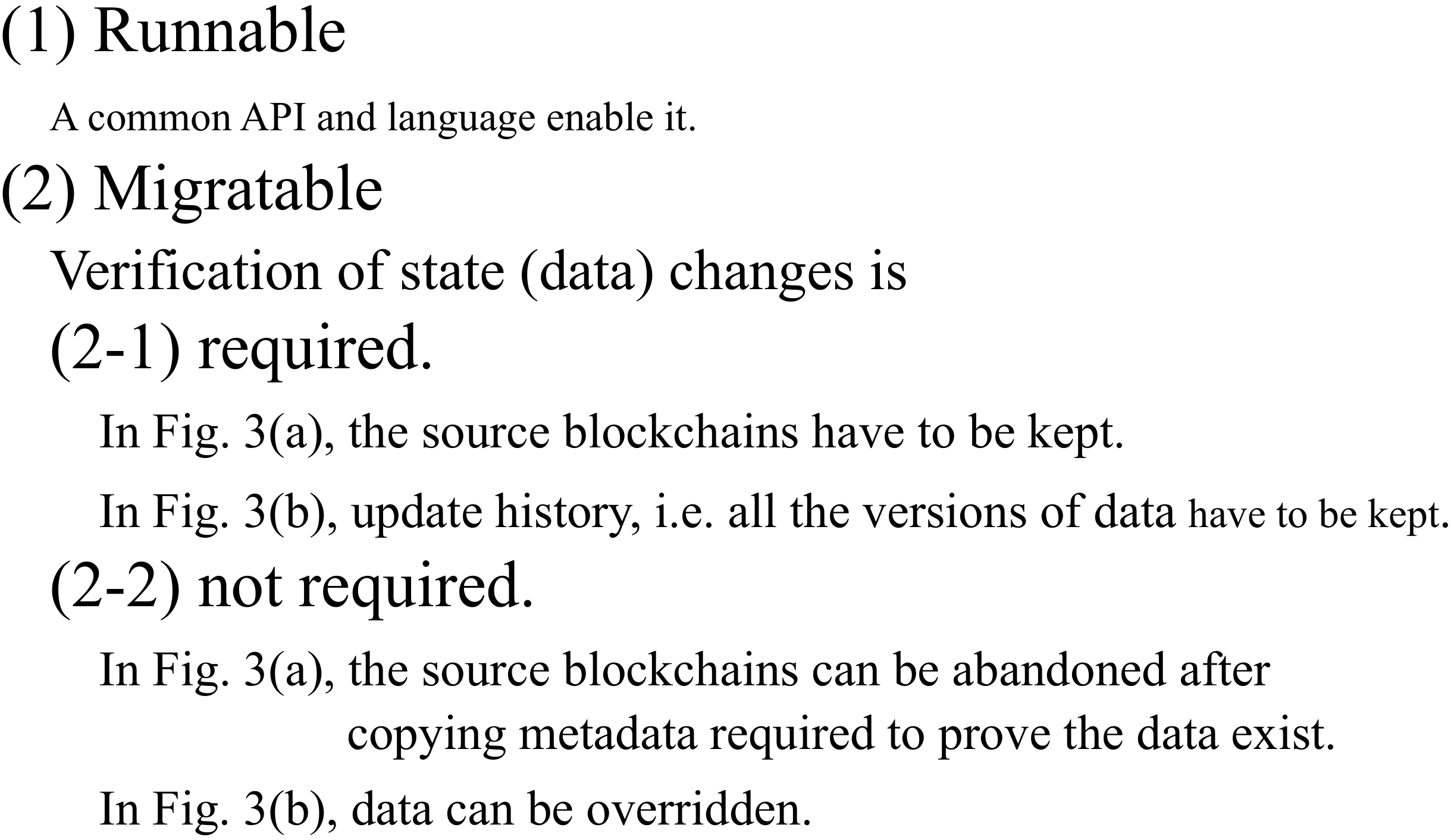}
  \caption{Design choices for blockchains supporting application migration.}
  \label{fig:design-choices}
\end{figure}
\else
%
% Figure 4
\begin{figure}[t]
  \centering
  \includegraphics[width=.48\textwidth,clip]{design-choices-fig2.pdf}
  \caption{Design choices for blockchains supporting application migration.}
  \label{fig:design-choices}
\end{figure}
\fi

Figure \ref{fig:design-choices} summarizes the design choices for blockchains supporting application migration.

\subsection{Migration process}

If an application requires access to all the logs (Section \ref{sec:mig-data}) in the original (source) blockchain, the middleware must provide such access.
However, we do not expect to be able to continue trusting the source blockchain (Section \ref{sec:arch}), for security reasons.
Furthermore, it is safer not to rely on the source blockchain's middlewares to still be running and accessible online.
In any case, we cannot generally control public blockchains, and maintaining private blockchain middleware requires human effort and CPU resources.
Thus, we need a way of maintaining access to previous blockchain logs without requiring the original blockchain to continue to be trustworthy, or even exist.

\ifINFLATE
%
% Figure 5
\begin{figure*}[t]
  \centering
  \includegraphics[width=.70\textwidth,clip]{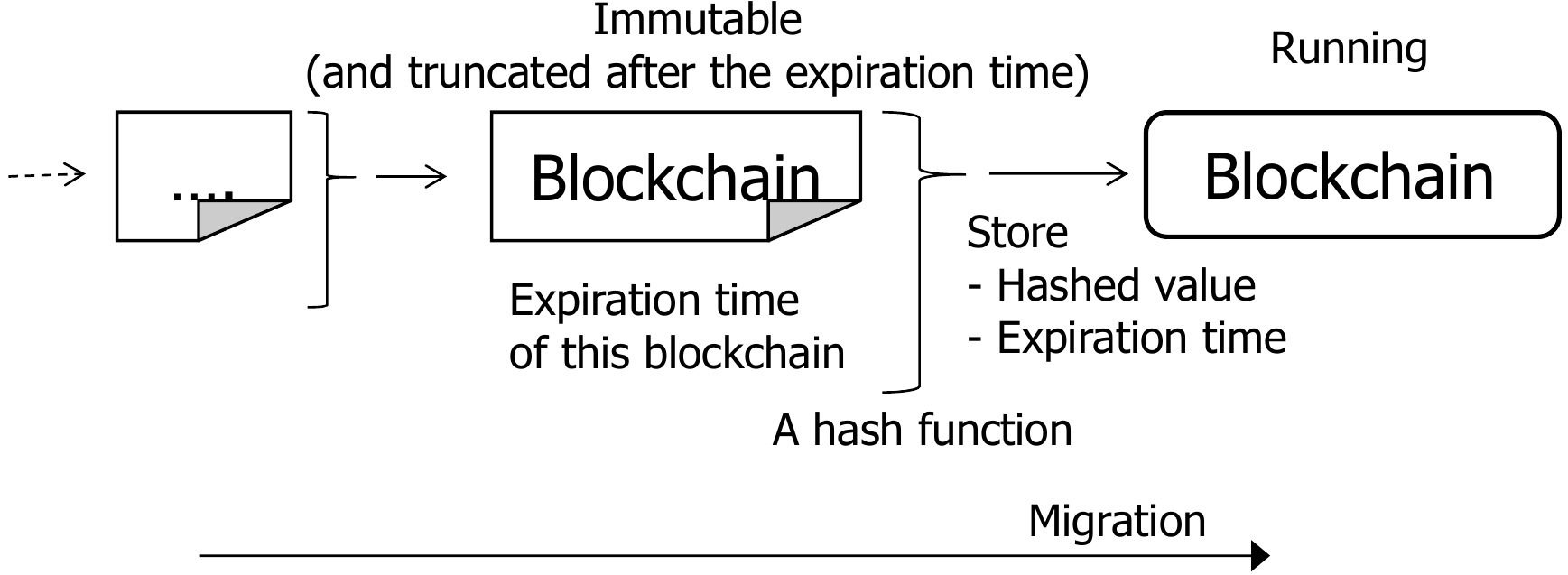}
  \caption{Technique for migrating an application between blockchains.}
  \label{fig:mig-blockchains}
\end{figure*}
\fi

\ifINFLATE
\else
%
% Figure 5
\begin{figure}[t]
  \centering
  \includegraphics[width=.49\textwidth,clip]{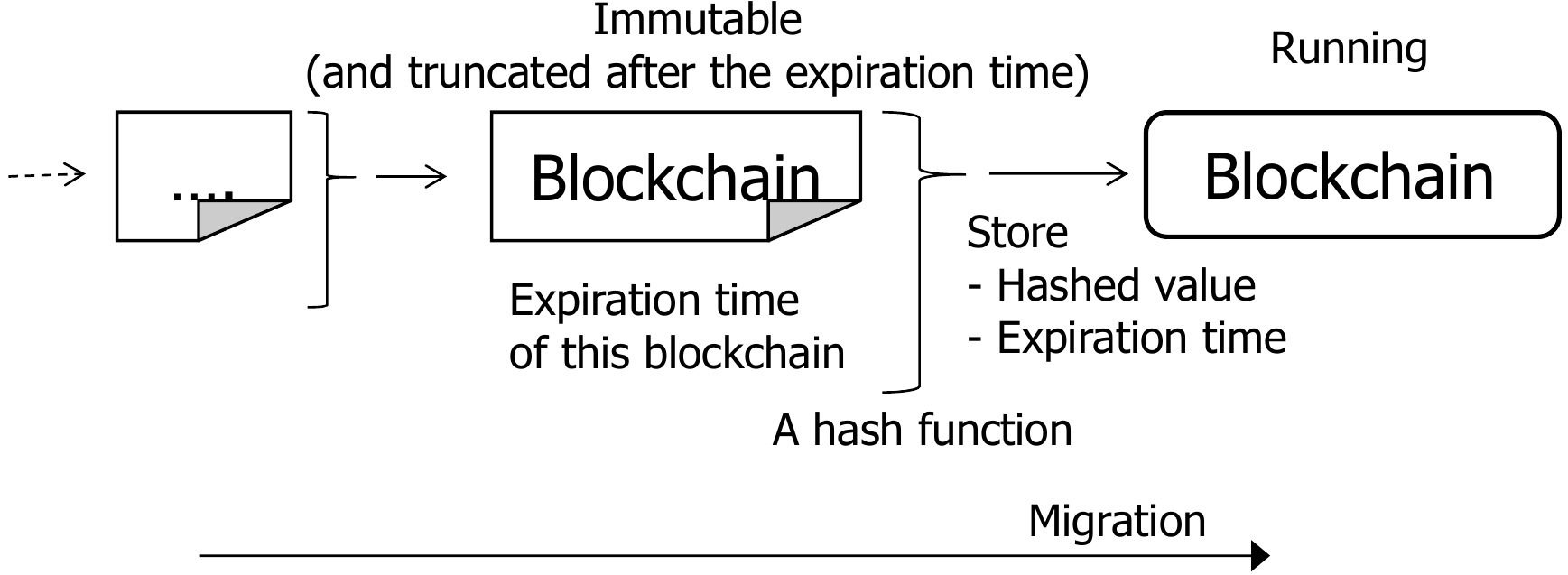}
%
%  \vspace{-1ex}
%
  \caption{Technique for migrating an application between blockchains.}
%
%  \vspace{-2ex}
%
  \label{fig:mig-blockchains}
\end{figure}
\fi

Figure \ref{fig:mig-blockchains} shows a migration process that meets these requirements.
The middleware stores a static copy of the source blockchain, truncating it at the expiration time, and accesses the stored blockchain.
Here, the middleware sets the source blockchain's expiration time to just before the migration.
The expiration time would be specified in block height because real-time time stamps can be manipulated to some degree.
Saving a static copy obviates the need for the blockchain middleware to continue running, and truncating it allows us to stop trusting the blockchain after the expiration time.

Then, the middleware creates a hash chain of the saved blockchains as shown in Figure \ref{fig:mig-blockchains}.
This hash chain enables it to detect any alterations of the saved blockchains, thus, proving it has not been modified, and confirm the stored expiration time is correct.

\ifINFLATE
%
% Figure 6
\begin{figure}[t]
  \centering
  \includegraphics[width=.49\textwidth,clip]{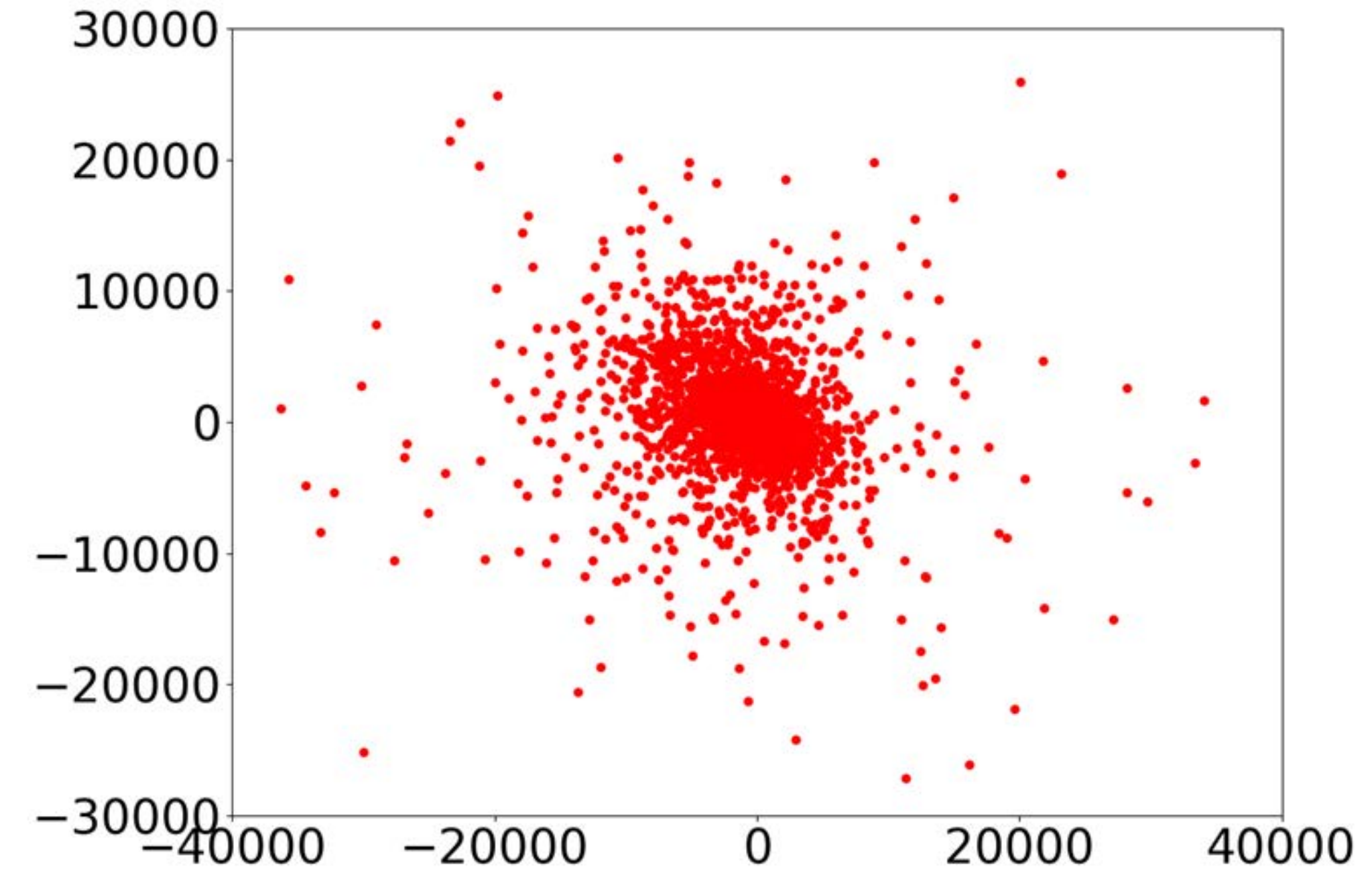}
  \caption{A network coordinate representing about 10,000 Bitcoin nodes (in millisecond).}
  \label{fig:net-coord}
\end{figure}
\fi

\ifINFLATE
Here we have to remember that it is possible for a once-confirmed block to be voided in a public blockchain.
It also happens naturally as well as by attacks mentioned in Section \ref{sec:motiv-pub}.
It seems that an application on the blockchain can just ignore such invalidation, because the application relies only on the saved blockchains, not still-running old blockchains.
Though there may be cases in which the application has to take account of the block invalidation.
In those cases the statically saved blockchains are also to be updated.
At least, expiration time of the saved blockchain has to be rewinded to a still-valid block.
Anyway, we want to avoid such updates, that are time-consuming and halt the application for the updates.
We are trying to reduce block propagation latency among nodes, that is the reason of naturally-happening fork and resulting block invalidation.
The techniques for reducing would be neighbor selection or route selection \cite{MiNS13}.
Such neighbor and route selections for performance requires network proximity information between nodes.
Therefore, first, we are trying to establish a technique to estimate communication latencies between nodes using network coordinates such as Vivaldi \cite{DCKM04}.
Figure \ref{fig:net-coord} shows communication latencies among Bitcoin nodes estimated using Vivaldi on a two-dimensional Euclidean plane.
The next step in this activity is accuracy improvement.
\fi

The middleware also has to be able to interpret all possible source blockchains, which may be in a variety of formats.
However, developing such interpreting functions takes significant effort, and the resulting software is likely to be complicated and bug-prone.
Therefore, we would prefer to simply abandon the source blockchains if the application allows it.
For applications that need to prove data exists but not verify state changes, we can abandon the source blockchain after copying the metadata needed to prove the data exist (and the data themselves in the case of Figure \ref{fig:arch} (a)) to the new blockchain.
If we take this approach, though, how can we prove the copied metadata are correct without the original metadata?
The trust provided by blockchains is based on verifiable data; however, such a copying approach means the copied metadata are not verifiable because they are not explicitly connected to the original metadata.
Although the application trusts the middleware that performed the copy, trust based on software correctness is weaker than trust gained through verifiable data because, for example, the middleware is subject to human mistakes.
The next-best solution to this problem would be to make the copying program verifiable, possibly by saving both it and its hashed value somewhere.

%It is not obvious how to migrate an application even between blockchains operated with the same middleware.

% todo: 以下，書き換える．

%It is relatively easier to migrate an application between different blockchains operated with the same middleware.
%It is not necessary to modify the application to migrate.
%But, even in the case, you have to migrate data in the source blockchain to the migration target.
%There is no standard way to do it, but at least it is possible to migrate such data by writing a program code, that reads state variables and writes them into the target in case of Ethereum.

% todo: 困難さを言いたいのだが… こんな記述しかできない?

\section{Conclusion}

%Although blockchain applications, like any other application, can benefit from being portable, those 
Applications for public blockchains are subject to the particular issue of incentive mismatch between the applications and the blockchain nodes.
This is because public blockchains are driven by different economic incentives than the applications that run on them, and are not under the applications' control.
Because of this, an application may also fail if its underlying blockchain collapses, unless it is not just portable but migratable.

These blockchain application survivability issues have motivated us to investigate application portability and migration.
In this paper, we have presented middleware designs that facilitate application migration between blockchains, together with more efficient alternative designs for applications with fewer requirements.

Making applications portable and migratable is both generally desirable and a potential solution to the incentive mismatch problem with public blockchains.
It would be better still for the blockchain to be supported by the applications themselves, not the nodes' economic incentives.
We can call it an {\em incentive-matched blockchain} if the public blockchain is always supported solely by the applications.
However, how to design such a mechanism remains an open problem.

Bodies such as the ISO have begun standardizing distributed ledgers, and the IETF and W3C have discussed such matters as well.
However, it is challenging just to design an effective common blockchain API and migration between blockchains lies even further in the future.
Today, software is implemented first and its specification is written after it gains much popularity.
It is implementation-first, not specification-first.
For instance, the first Bitcoin Improvement Proposal (BIP) came more than two and a half years after the Bitcoin network was launched.
Following this paper, our effective next step will be to demonstrate applications that are portable and migratable.
%Anyway, we will have a standardized terminology and concepts and they are an important base for portability.

In future work, we plan to study the portability of program code and data for smart contracts, a topic this paper has been barely touched on.
Even on the same middleware platform, it is difficult to update the data format or program code while retaining existing data.

\section*{Acknowledgments}

We thank Yuto Takei, Masashi Hojo, and Shigeya Suzuki for discussions on a number of topics, such as the incentive mismatch problem and future incentive-matched public blockchains.
We also thank Toshio Koide for
pointing out the significance of portability on private blockchains.
%the discussion on portability of smart contracts.
%
This work was supported by Kaula, Inc.,
the SECOM Science and Technology Foundation,
the New Energy and Industrial Technology Development Organization (NEDO),
and JSPS KAKENHI Grant Numbers 2570008 and 16K12406.

\bibliography{blockchain}
\ifIEEE
\bibliographystyle{unsrt}
\fi
\ifACM
\bibliographystyle{abbrv}
\fi

\end{document}